\pdfoutput=1

\documentclass[aps,pra,twocolumn,superscriptaddress]{revtex4}

\usepackage{amssymb,amsmath,graphicx}

\begin{document}

\title[Exact Solution of the Dynamo Problem]%
{Exact Solution of the Direct and Inverse Quasi-stationary
Dynamo Problem \\ in the Expanding Plasma Ball}

\author{Yurii V. Dumin}
\email[Electronic mail: ]{dumin@pks.mpg.de, dumin@yahoo.com}
\affiliation{Lomonosov Moscow State University,
Sternberg Astronomical Institute, \\
Universitetskii prosp.\ 13, 119234 Moscow, Russia}
\affiliation{Space Research Institute of Russian Academy of Sciences, \\
Profsoyuznaya str.\ 84/32, 117997 Moscow, Russia}
\affiliation{\rm ORCID: 0000-0002-9604-5304}

\date{22 October 2026}

\begin{abstract}
It is shown that differential equation of the quasi-stationary dynamo
effect (\textit{i.e.}, generation of the electric fields and currents) in
a uniformly-expanding plasma ball with strongly anisotropic conductivity
possesses the unique mathematical property:
namely, the spectrum of its eigenvalues is universal and independent of
physical parameters of the medium.
As a result, it becomes possible to introduce a special set of
eigenfunctions---which we called the generalized spherical functions---that
can be used to solve the dynamo problem in exactly the same way as ordinary
spherical functions are used to solve the Laplace equation.
The corresponding exact solutions should be especially valuable for treating
the inverse dynamo problem, \textit{i.e.}, determination of the plasma
parameters from the experimentally measured electric fields and currents.
\\
\textbf{Keywords:} dynamo-effect, anisotropic plasma, inverse problem
\end{abstract}

\maketitle

\section{Introduction}
\label{sec:Intro}

The problem of generation of the electric fields and currents by
an expanding plasma ball is of importance in various branches of physics.
For example, the so-called active space experiments---where artificial
plasma clouds are released from rockets and satellites into the near-Earth
space---are widely conducted since 1960s~\cite{Haerendel_68,Marklund_87}.
Their aim was both to study the electrodynamic effects produced when
the expanding cloud interacts with the environmental plasmas as well as
to extract the plasma parameters (conductivity, concentration, collisional
frequencies, \textit{etc.}) from the measured electric fields and currents
\textit{i.e.}, to solve the inverse dynamo problem.
(To avoid misunderstanding, let us emphasize that this definition of
the dynamo effect is irrelevant to the problem of generation or alteration
of the magnetic field itself, which can take place both in the active space
experiments~\cite{Gavrilov_99} and other astrophysical
situations~\cite{Zeldovich_83}.)

Later, since the early 2000s, a quite similar problem appeared in
the laboratory experiments with the so-called ultracold plasmas (UCP),
which were created in the magneto--optical traps and then experienced a free
expansion in space~\cite{Killian_99,Killian_07b}.
Such experiments were initially performed without any magnetic fields, but
subsequently a considerable attention was paid also to the dynamics of UCP
embedded into the external magnetic field, either
uniform~\cite{Zhang_08,Sprenkle_22} or possessing a more complex
configuration~\cite{Bronin_23}.
Solution of the inverse dynamo problem in this situation would be a valuable
tool, particularly, for diagnostics of the electron component of UCP.
Indeed, while the ions can be well detected by the laser-induced fluorescence,
the ability to diagnose electrons remains much more limited.
This is actually performed only by counting the number of particles incident
onto the position-sensitive detectors.
So, deriving the components of the electric conductivity tensor from
the solution of inverse dynamo problem should give a lot of additional
information, \textit{e.g.}, about the effective collisional frequencies,
\textit{etc.}

In fact, a number of attempts to solve the inverse dynamo problem were
undertaken quite a long time ago in the context of active space experiments;
see paper~\cite{Marklund_87} and references therein.
Unfortunately, they were carried out under the very strong
simplifications---typically, in the 2D~approximation and only for integral
(along the magnetic field lines) electric conductivities.
So, the aim of our work is to present the method of constructing the exact
3D~solutions.
This method is based on the unique mathematical property of the dynamo-effect
differential equation, which was unexpectedly found in the course of our
analysis.
Namely, the spectrum of eigenvalues of the above-mentioned equation is
``universal'' (\textit{i.e.}, independent of the physical parameters of
the medium) and, moreover, coincides with the spectrum of Legendre polynomials.
As a result, it becomes possible to introduce the special set of basic
functions---which will be called the generalized spherical functions---that
can be used to solve the dynamo equation in exactly the same way as ordinary
spherical functions (Legendre polynomials) are used for solving the Laplace
equation~\cite{Mathews_64,Arfken_70}.
(Of course, the generalized spherical functions themselves are not
``universal'', \textit{i.e.}, depend substantially on the particular
physical parameters of the medium.)

\section{Formulation of the Problem}
\label{sec:Formulation}

We shall formulate the dynamo problem in the quasi-stationary approximation,
when time derivatives of all physical quantities are ignored (for more details
about the conditions of its applicability, see paper~\cite{Marklund_87}).
Then, a continuity equation for the electric current can be written just as
\begin{equation}
\nabla{\cdot}\mathbf{j} = 0 \, .
\label{eq:cont_eq}
\end{equation}

\begin{figure}
\includegraphics[width=0.95\columnwidth]{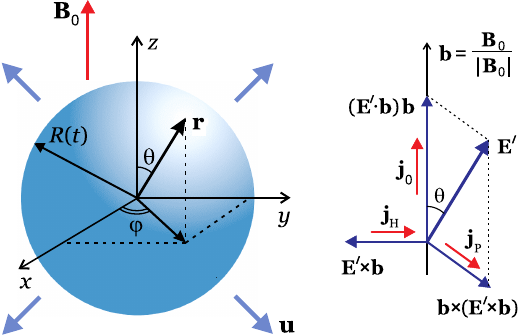}
\caption{\label{fig:Ball_and_Electr_current}
Sketch of the expanding plasma ball (left panel) and
decomposition of the total current density into
the longitudinal~$ {\bf j}_0 $, Pedersen~$ {\bf j}_{\rm P} $, and
Hall~$ {\bf j}_{\rm H} $ currents (right panel).}
\end{figure}

Next, the electric current density~$ \textbf{j} $ in the anisotropic plasma
can be expressed in terms of the electric field~$ \textbf{E}^\prime $ in
the co-moving reference frame by the generalized Ohm law:
\begin{equation}
\mathbf{j} = \sigma_0 (\mathbf{E}^\prime{\cdot}\mathbf{b}) \mathbf{b} +
\sigma_{\rm P} (\mathbf{b}{\times}(\mathbf{E}^\prime{\times}\mathbf{b})) -
\sigma_{\rm H} (\mathbf{E}^\prime{\times}\mathbf{b}) \, ,
\label{eq:Ohm_law}
\end{equation}
where $ \sigma_0 $, $ \sigma_{\rm P} $, and $ \sigma_{\rm H} $~are
the longitudinal, Pedersen, and Hall conductivities (their explicit
expressions in terms of the plasma parameters can be found, \textit{e.g.},
in monograph~\cite{Bostrom_73}); while
$ \textbf{b} = \textbf{B}_0 / |\textbf{B}_0| $ is the unitary vector in
the direction of the external magnetic field~$ \textbf{B}_0 $;
see right panel in Fig.~\ref{fig:Ball_and_Electr_current}.

The electric field~$ \textbf{E}^\prime $ in the co-moving reference frame can
be reduced to the one in the laboratory frame~$ \textbf{E} $ by the standard
transformation:
\begin{equation}
\mathbf{E}^\prime = \mathbf{E} + \mathbf{u}{\times}\mathbf{B}_0 / c \, ,
\label{eq:E_transform}
\end{equation}
where $ \textbf{u} $~is the macroscopic plasma velocity, and
$ c $~is the speed of light.
At last, since the time derivatives are everywhere ignored, the electric
field~$ \textbf{E} $ can be treated in the electrostatic approximation and,
therefore, expressed through the electric potential~$ \Phi $:
\begin{equation}
\mathbf{E} = -\nabla{\Phi} \, .
\label{eq:electrostat_pot}
\end{equation}

From here on, we shall assume for simplicity that
(i)~the plasma cloud is spherically symmetric and uniform, \textit{i.e.},
its parameters, such as $ \sigma_0 $, $ \sigma_{\rm P} $,
and~$ \sigma_{\rm H} $, are constant in space but can depend on
the time~$ t $ as a parameter, and
(ii)~the external magnetic field~$ \textbf{B}_0 $ is constant in time and
uniform in space, \textit{i.e.}, unperturbed by the plasma motion.
So, it is convenient to use the spherical coordinate
system~$ (r, \vartheta, \varphi) $ with polar axis directed along
the magnetic field~$ \textbf{B}_0 $, as depicted in the left panel of
Fig.~\ref{fig:Ball_and_Electr_current}.

Next, a boundary of the plasma cloud will be assumed to move by
the linear law:
\begin{equation}
R(t) = R_0 + u_0 t \, .
\label{eq:ball_boundary}
\end{equation}
From the physical point of view, this corresponds to the inertial stage of
expansion, when the most part of initial thermal energy of the gas was
transformed into kinetic energy of its macroscopic motion.
The mass density~$ \rho $ of the uniform cloud should change with time as
\begin{equation}
\rho(t) =  \rho_0 \big( R_0 / R(t) \big)^{\! 3} =
\rho_0 \big( 1 + u_0 t / R_0 \big)^{\! -3} \, .
\label{eq:rho_time}
\end{equation}
Then, employing the mass continuity equation,
\begin{equation}
\partial \rho / \partial t + \nabla{\cdot}(\rho \mathbf{u}) = 0 \, ,
\label{eq:mass_contin}
\end{equation}
and assuming that the velocity field~$ \textbf{u} $ has only the radial
component and depends only on the time and radius, we easily find that
\begin{equation}
u_r(r, t) = \big[ u_0 / R(t) \big] \, r \, ,
\label{eq:u_r}
\end{equation}
\textit{i.e.}, the velocity should be a linear function of radius.

Finally, combining all the above-written equations and taking into account
that $ \partial\Phi / \partial\varphi \equiv 0 $, we get the second-order
partial differential equation defining a distribution of the electric
potential inside the expanding plasma ball:
\begin{eqnarray}
&& \!\!\!\!\!\!\!
(\sigma_0\,{-}\,\sigma_{\rm P}) \big[
  \xi^2 r^2 \Phi^{\prime\prime}
  + 2(1{-}\xi^2) \xi r \dot{\Phi}^{\prime}
  + (1{-}\xi^2)^2 \ddot{\Phi}
\nonumber \\
&& \qquad \qquad \qquad \quad \:
  + \: (1{-}\xi^2) r \Phi^{\prime}
  - 3 (1{-}\xi^2) \xi \dot{\Phi} \big]
\nonumber \\
&& \quad \!
+ \: \sigma_{\rm P} \big[ r^2 \Phi^{\prime\prime} +
  (1{-}\xi^2) \ddot{\Phi}
  + 2 r \Phi^{\prime} - 2 \xi \dot{\Phi} \big] =
\nonumber \\
&& \qquad \qquad \qquad \qquad \qquad \qquad
2 \sigma_{\rm H} \big( \! B_0 u_0 / c R \big) r^2 \! ,
\label{eq:pot_distrib}
\end{eqnarray}
where $ \xi = \cos{\vartheta} $;
the prime denotes a differentiation with respect to~$ r $; and
dot, with respect to~$ \xi $.
In the case of isotropic medium ($ \sigma_0\,{=}\,\sigma_{\rm P} $),
the left-hand side of the above-written expression is evidently reduced
to the Laplace equation in spherical coordinates.

Let us briefly discuss the scope of applicability of the assumptions
used to derive the dynamo-equation~(\ref{eq:pot_distrib}).
A crucial conjecture was, of course, the uniform distribution of density
within a sharply-bounded cloud.
In fact, the plasma clouds---both in the active space experiments and
the laboratory UCP installations---typically have the smooth Gaussian-type
density profiles.
However, since the Gaussian function has the characteristic length scale
(r.m.s.\ deviation), just this quantity could be reasonably associated with
radius of the uniform cloud.
Moreover, as follows from the numerous experiments~\cite{Killian_07b},
the above-mentioned length scale increases linearly with time, which is
in agreement with our conjecture~(\ref{eq:ball_boundary}).
Therefore, the employed model of a ``rectangular'' density profile is crude
but meaningful.
Moreover, as follows from the recent modelling~\cite{Vikhrov_20}, expansion
of the UCP cloud can lead to the formation of the ionic front.
This gives an additional argument in favour of the sufficiently sharp
boundary.

At last, let us remind once again that the final
equation~(\ref{eq:pot_distrib}) was derived in the quasi-stationary
approximation, where time serves merely as a parameter.
In the more general case, formula~(\ref{eq:cont_eq}) should be replaced by
the continuity equation for the electric charge:
\begin{equation}
\frac{\partial {\rho}_e}{\partial t} + \nabla{\cdot}\mathbf{j} = 0 \, ,
\label{eq:cont_eq_gen}
\end{equation}
where $ {\rho}_e $~is the electric charge density.
Let $ L $ and $ T $~be the characteristic (macroscopic) spatial and temporal
scales of the dynamo-problem.
Then, the first term in the left-hand side of Eq.~(\ref{eq:cont_eq_gen}) can
be estimated as:
$ {\partial {\rho}_e}/{\partial t} \sim {{\rho}_e}/{T} \sim E/LT $
(taking into account that $ \nabla{\cdot}\mathbf{E} = 4 \pi {\rho}_e $).
On the other hand, the second term is estimated as:
$ \nabla{\cdot}\mathbf{j} \sim j/L \sim {\sigma}E/L $
(where $ \sigma $~is the typical conductivity).
Therefore, the first term can be ignored if
$ \sigma T \gg 1 $.
If this is not the case, both terms should be taken into account; so that
the resulting dynamo-problem will be described by a time-dependent
equation of parabolic type.

In addition, since the time derivatives are taken into consideration,
expression~(\ref{eq:electrostat_pot}) for the electric field intensity
should be replaced by
$
\mathbf{E} = -\nabla{\Phi} - (1/c)\,{\partial \mathbf{A}}/{\partial t} ,
$
where $ \mathbf{A} $~is the vector potential, which is related to
perturbation of the magnetic field due to the local electric currents:
$
\delta \mathbf{B} = \nabla{\times}\mathbf{A}
$.
In other words, the magnetic field becomes inherently non-stationary.
Anyway, treating the dynamo effect in this more general formulation is
much more challenging task, which is far beyond the scope of the present
paper.

\section{The Method of Solution}
\label{sec:Method}

Let us seek a solution of the equation~(\ref{eq:pot_distrib}) as a power
series of radius:
\begin{equation}
\Phi(r, \xi) = \sum\limits_{n = 0}^\infty f_n(\xi) \, r^n ,
\label{eq:series_r}
\end{equation}
where $ f_n(\xi) $~are the unknown functions, and
the terms with $ n < 0 $ were ignored to avoid singularity in the origin
of coordinates.

Then, substituting expansion~(\ref{eq:series_r}) into
equation~(\ref{eq:pot_distrib}) and collecting the terms with the same
powers of~$ r $, we get:
\begin{eqnarray}
&&
A \, \big\{ \! (1{-}\xi^2)^2 \ddot{f}_n
  + (2n - 3) (1{-}\xi^2) \xi \dot{f}_n
\nonumber \\[-0.2ex]
&& \qquad \qquad \qquad
  + \, n \big[ (n - 2) \xi^2 + 1 \big] f_n \! \big\}
\nonumber \\
&&
\;\; + \: \big\{ \! (1{-}\xi^2) \ddot{f}_n
  - 2 \xi \dot{f}_n
  + n (n + 1) f_n  \! \big\} =
\nonumber \\[0.6ex]
&& \qquad \qquad \qquad \qquad
2 \, \big( \sigma_{\rm H} / \sigma_{\rm P} \big)
  \big( B_0 u_0 / c R \big) \, \delta_{n2} \, ,
\label{eq:f_n}
\end{eqnarray}
where $ n = 0, 1, \dots $;
$ A $~is the parameter of anisotropy, defined as
\begin{equation}
A = ( \sigma_0\,{-}\,\sigma_{\rm P} ) / \sigma_{\rm P} \, ;
\label{eq:A_def}
\end{equation}
and $ \delta $~is the Kronecker delta symbol
($ \delta_{n2} = 0 $ at $ n \neq 2 $, and $ \delta_{22} = 1 $).
Let us emphasize that each equation~(\ref{eq:A_def}) contains the unknown
functions~$ f_n $ with the same value of~$ n $.
Therefore, we do not need to solve an infinite sequence of interlinked
equations.

If $ n \neq 2 $, the ordinary differential equations~(\ref{eq:A_def}) are
homogeneous.
On the other hand, the equation at $ n = 2 $ is inhomogeneous.
So, its general solution can be presented as a sum of any particular
solution and a general solution of the respective homogeneous equation.
Next, let us note that the coefficient of undifferentiated function~$ f_2 $
in the left-hand side of equation~(\ref{eq:A_def}) does not depend on~$ \xi $.
So, a particular solution (which will be marked by asterisk) can be taken
just as the constant:
\begin{equation}
f^*_2 = \big[ \! B_0 u_0 / (c R) \big]
  \big( \sigma_{\rm H} / \sigma_{\rm P} \big) / (A + 3) \, .
\label{eq:part_sol}
\end{equation}

Therefore, general solution~(\ref{eq:series_r}) of the original
equation~(\ref{eq:pot_distrib}) can be rewritten as
\begin{eqnarray}
\Phi(r, \xi) & = &
  \frac{\displaystyle B_0 u_0 R}{c} \,
    \frac{(\sigma_{\rm H} / \sigma_{\rm P})}{A + 3} \,
    \bigg( \! \frac{r}{R} \! \bigg)^{\!\! 2}
\nonumber \\
  & + & \sum\limits_{n = 0}^\infty C_n F_n(\xi) \,
    \bigg( \! \frac{r}{R} \! \bigg)^{\!\! n} ,
\label{eq:gen_sol}
\end{eqnarray}
where coefficients~$ C_n $ should be determined by the boundary conditions
on the sphere $ r = R $.

Next, to determine the ``generalized'' spherical functions~$ F_n(\xi) $,
we perform again their expansion into the power series:
\begin{equation}
F_n(\xi) = \sum\limits_{k = 0}^\infty a_{nk} \xi^k ,
\label{eq:gen_sper_func}
\end{equation}
where negative powers ($ k < 0 $) were omitted to avoid singularity
in the equatorial plane ($ \vartheta = \pi / 2 $, $ \xi = 0 $).
Substituting the above-written expansions into the homogeneous
equations~(\ref{eq:pot_distrib}) and collecting the terms with the same
powers of~$ \xi $, we get the following set of algebraic recursion relations
for three successive coefficients~$ a_{nk} $ with the same parity:
\begin{eqnarray}
(1 - \delta_{k0} - \delta_{k1}) A (n - k) (n - k + 2) \! & \!\! a_{n, k-2}
\nonumber \\
+ \, (n - k)[(2k + 1)A + (n + k + 1)] \! & \!\!\!\!\!\!\!\!\! a_{nk}
\nonumber \\
+ \, (k + 1)(k + 2)(A + 1) \! & \!\! a_{n, k+2} & \!\! = 0 \: ,
\label{eq:recur_rel}
\end{eqnarray}
where $ \delta_{k0} $ and $ \delta_{k1} $~are the Kronecker delta symbols.

At the first sight, at the fixed~$ n $ and increasing indices~$ k $
the above-written formulas represent an infinite sequence of interlinked
relations, and each of them (for odd and even~$ k $, respectively) is
determined by the first two terms.
However, a more careful analysis shows that this is not the case:
each of these sequences contains only a finite number of terms and is
completely determined by a single coefficient.
Really, at $ k = 0 $ or~1 the first term in formula~(\ref{eq:recur_rel})
vanishes, so that the third coefficient ($ a_{n2} $ or $ a_{n3} $) is
uniquely determined if the second coefficient ($ a_{n0} $ or $ a_{n1} $)
was specified.
Next, all subsequent coefficients~$ a_{nk} $ with the same parity can be
easily derived from the three-term recursion relations~(\ref{eq:recur_rel})
since two preceding coefficients are known.
Without a loss of generality, the first non-zero coefficient in each sequence
can be equated to unity (\textit{i.e.}, $ a_{n0} = 1 $ for even~$ n $ and
$ a_{n1} = 1 $ for odd~$ n $), because this simply implies a redefinition of
the coefficients~$ C_n $ in the expansion~(\ref{eq:gen_sol}).

Furthermore, if the recursion relations~(\ref{eq:recur_rel}) are written for
even~$ k $ at the even~$ n $ and odd~$ k $ at the odd~$ n $, it can be shown
that they are interrupted at $ k = n $ (\textit{i.e.}, $ a_{kn} \equiv 0 $
at $ k > n $).
Really, let us take $ k = n $ in formula~(\ref{eq:recur_rel}).
Then, the first two terms vanish due to the multipliers~$ (n - k) $, while
the third term remains non-zero.
Consequently, $ a_{n,k+2} = 0 $.
Next, if we take~$ k = n + 2 $, then the first term in
formula~(\ref{eq:recur_rel}) vanishes due to the multiplier~$ n - k + 2 $,
and the second term is equal to zero because $ a_{n,k+2} = 0 $.
Therefore, $ a_{n,k+4} = 0 $.
Finally, if two successive coefficients with the same parity in the three-term
recursion relations are equal to zero, all subsequent coefficients will
evidently also vanish.
This truncation criterion turns out to be universal, \textit{i.e.},
independent of the arbitrary real parameter~$ A $.
Therefore, the generalized spherical functions~$ F_n(\xi) $ are polynomials of
a finite order, whose algebraic structure is exactly the same as for
the well-known Legendre polynomials~\cite{Arfken_70,Mathews_64} (although
their coefficients~$ a_{nk} $ are much more complex and depend substantially
on the anisotropy parameter~$ A $).

Summarizing the preceding consideration, we can write general solution of
the dynamo problem in the following compact form:
\begin{eqnarray}
\Phi(r, \xi) & \! = \! & \frac{B_0 u_0 R}{c}
  \bigg( \! \frac{\sigma_{\rm H}}{\sigma_{\rm P}} \! \bigg)
  \bigg[ \frac{1}{A+3} \bigg( \! \frac{r}{R} \! \bigg)^{\!\! 2}
\nonumber \\
  & \! + \! & \sum\limits_{k = 0}^\infty \xi^k
  \sum\limits_{n = k}^\infty
  C_n a_{nk} \bigg( \! \frac{r}{R} \! \bigg)^{\!\!\! n} \bigg] ,
\label{eq:gen_sol_compact}
\end{eqnarray}
where we changed the summation order and took into account that
$ a_{kn} \equiv 0 $ at $ k > n $.
Besides, the arbitrary coefficients~$ C_n $ were redefined, as compared to
formula~(\ref{eq:gen_sol}), to make them dimensionless.
Next, these coefficients should be specified for the particular boundary
conditions.

\begin{figure}
\includegraphics[width=0.75\columnwidth]{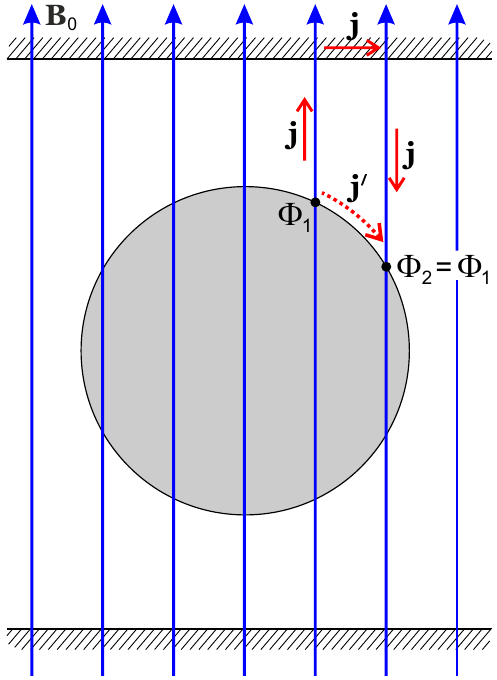}
\caption{\label{fig:Short_circuit}
Sketch of the short-circuit effect, occurring when the external plasma is
strongly magnetized (\textit{i.e.}, anisotropic) but surrounded by
the perfectly conducting (isotropic) walls.}
\end{figure}

For example, if we assume a perfectly-conducting external medium, then
a tangential component of the electric field should vanish at the ball
surface:
\begin{equation}
\frac{\partial \Phi_c(r, \xi)}{\partial \xi} \bigg|_{r = R} \! = 0 \, .
\Bigg.
\label{eq:BC_conduct}
\end{equation}
The corresponding physical quantities will be marked by subscript~`c'
(conducting).
In fact, this condition is applicable not only when the external medium is
perfectly conducting in all directions.
Yet another important case is when the outer plasma is strongly magnetized
(\textit{i.e.}, its conductivity is high only along the magnetic field and
tends to zero in two perpendicular directions), but the magnetic field lines
intersect the conducting walls of the experimental chamber, thereby resulting
in the short-circuit effect; see Fig.~\ref{fig:Short_circuit}.
Then, the above-written condition~(\ref{eq:BC_conduct}) will formally
remain valid.

Substituting the general solution~(\ref{eq:gen_sol_compact}) into
formula~(\ref{eq:BC_conduct}) and collecting the terms with the same powers
of~$ \xi $, we get a set of linear algebraic equations for
the coefficients~$ C_n $:
\begin{equation}
\sum\limits_{n = k}^{\infty} a_{nk} C_n = 0 \, , \quad
k = 1, 2, \dots, \infty \, .
\label{eq:BC_conduct_eqs}
\end{equation}
Since matrix~$ |\!| a_{nk} |\!| $ is triangular, then
\begin{equation}
\det |\!| a_{nk} |\!| =
\prod\limits_{n = 1}^{\infty} a_{nn} \neq  0 \, ,
\label{eq:det_a}
\end{equation}
\text{i.e.}, the system of homogeneous linear
equations~(\ref{eq:BC_conduct_eqs}) is non-degenerate.
Therefore, it has only the trivial solution:
\begin{equation}
C_n = 0 \, , \quad n = 1, 2, \dots, \infty \, .
\label{eq:BC_conduct_sol}
\end{equation}
The value of~$ C_0 $ remains indefinite, which corresponds to
the arbitrary additive constant in the potential, \textit{e.g.},
$ C_0 = 0 $.
As a result, solution for the perfectly-conducting external
medium is reduced just to the first term of the general
expression~(\ref{eq:gen_sol_compact}):
\begin{equation}
\Phi_c(r) = \frac{B_0 u_0 R}{c}
  \frac{1}{A+3} \bigg( \! \frac{\sigma_{\rm H}}{\sigma_{\rm P}} \! \bigg)
   \bigg( \! \frac{r}{R} \! \bigg)^{\!\! 2} .
\label{eq:part_sol_cond}
\end{equation}

Following the same procedure, one can find solutions for any other boundary
conditions.
For example, if the plasma cloud is surrounded by isolating (dielectric)
external medium (\textit{e.g.}, vacuum), a normal component of the electric
current should vanish on the ball surface.
Using expressions~(\ref{eq:Ohm_law})--(\ref{eq:electrostat_pot}), this can
be formulated as
\begin{eqnarray}
(1 + A \xi^2) \, \frac{\partial \Phi_d(r, \xi)}{\partial r} \bigg|_{r = R} \!
  & \! + \! & \!\! \bigg( \! \frac{A}{R} \! \bigg) (1 - \xi^2) \, \xi \,
  \frac{\partial \Phi_d(r, \xi)}{\partial \xi} \bigg|_{r = R}
\nonumber \\[0.5ex]
  & \! = \! & \frac{B_0 u_0}{c}
  \bigg( \! \frac{\sigma_{\rm H}}{\sigma_{\rm P}} \! \bigg) (1 - \xi^2) \, ,
\label{eq:BC_isol}
\end{eqnarray}
where subscript~`d' implies dielectric.
Next, substituting the general expression~(\ref{eq:gen_sol_compact})
to~(\ref{eq:BC_isol}) and solving the system of linear algebraic equations
for the coefficients~$ C_n $, we get the required distribution of
the electric potential:
\begin{equation}
\Phi_d(r, \xi) = \frac{B_0 u_0 R}{2 c}
  \bigg( \! \frac{\sigma_{\rm H}}{\sigma_{\rm P}} \! \bigg)
  \bigg( \! \frac{r}{R} \! \bigg)^{\!\! 2} (1 - \xi^2) \, .
\label{eq:part_sol_isol_xi}
\end{equation}

\section{Results and Discussion}
\label{sec:Results}

\begin{figure*}[t]
\includegraphics[width=0.98\textwidth]{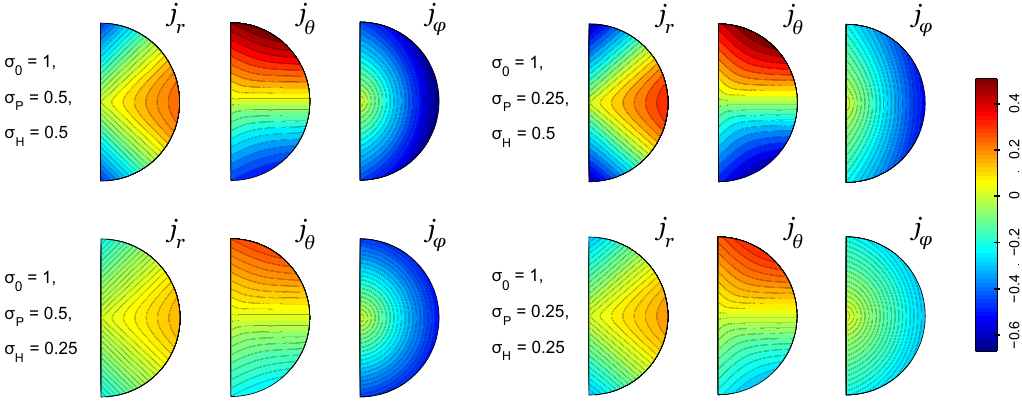}
\caption{\label{fig:Currents}
Examples of the current density distribution in the meridional
plane~$ (r, \vartheta) $ at various conductivities
(in arbitrary units).}
\end{figure*}

Let us return to the case of conducting boundary condition, which is most
interesting for the laboratory experiments.
Substituting the corresponding solution~(\ref{eq:part_sol_cond}) into
the formulas~(\ref{eq:Ohm_law})--(\ref{eq:electrostat_pot}), we get
expressions for the electric field,
\begin{eqnarray}
E^{\prime}_{cr} & \!\! = \!\! & - \frac{2 B_0 u_0}{c} \bigg[
  \bigg( \! \frac{{\sigma}_0}{{\sigma}_{\rm P}} \! \bigg) + 2 \, \bigg]^{-1}
  \bigg( \! \frac{{\sigma}_{\rm H}}{{\sigma}_{\rm P}} \! \bigg) \,
  \frac{r}{R} \, ,
\nonumber
\\
E^{\prime}_{c\varphi} & \!\! = \!\! & - \frac{u_0 B_0}{c} \, \frac{r}{R} \, ,
\label{eq:E_c}
\end{eqnarray}
and current,
\begin{eqnarray}
j_{cr} \!\! & = &
  \! {\sigma}_{\rm H} \, \frac{B_0 u_0}{c} \, \frac{r}{R} \,
  \bigg[ \! - \! \frac{2}{\tilde{\sigma}_0 + 2}
  \big( \tilde{\sigma}_0 \, {\cos}^2\vartheta + {\sin}^2\vartheta \big)
  + \sin\vartheta \bigg] ,
\nonumber
\\
j_{c\vartheta} \!\! & = \! &
  \! {\sigma}_{\rm H} \, \frac{B_0 u_0}{c} \, \frac{r}{R} \,
  \bigg[ \, \frac{2}{\tilde{\sigma}_0 + 2}
  \big( \tilde{\sigma}_0 - 1 \big) \sin\vartheta \cos\vartheta
  + \, \cos\vartheta \bigg] ,
\nonumber
\\
j_{c\varphi} \!\! & = &
  \! {\sigma}_{\rm H} \, \frac{B_0 u_0}{c} \, \frac{r}{R} \,
  \bigg[ \! - \! \frac{2}{\tilde{\sigma}_0 + 2} \,
  \frac{\sin{\vartheta}}{\tilde{\sigma}_{\rm P}} - \tilde{\sigma}_{\rm P}
  \bigg] .
\label{eq:j_cond}
\end{eqnarray}
Here, we introduced for conciseness the ``normalized'' electric
conductivities:
$ \tilde{\sigma}_0 = {\sigma}_0 / {\sigma}_{\rm P} $
and
$ \tilde{\sigma}_{\rm P} = {\sigma}_{\rm P} / {\sigma}_{\rm H} $.

As is known, the longitudinal conductivity~$ {\sigma}_0 $ is always greater
than the Pedersen and Hall conductivities, $ {\sigma}_{\rm P} $
and~$ {\sigma}_{\rm H} $; but a relationship between the two last-mentioned
quantities can be different, depending on the collisional and gyrofrequencies
of the electrons and ions~\cite{Bostrom_73}.
So, for illustration, let us consider the case when~$ {\sigma}_0 = 1 $
(in arbitrary units), while $ {\sigma}_{\rm P} $ and~$ {\sigma}_{\rm H} $
can take the values~0.5 and~0.25 in various combinations, depicted in
Fig.~\ref{fig:Currents}.
As can be seen, the electric current outflows from the plasma ball near
the equatorial plane and returns back near the poles; and its intensity
increases at higher values of the Hall conductivity~$ {\sigma}_{\rm H} $.
On the other hand, the azimuthal current~$ j_{c\varphi} $ is concentrated
in the outer parts of the ball.
Its distribution is represented by the approximately spherical shells at
$ {\sigma}_{\rm P} \geqslant {\sigma}_{\rm H} $ but becomes spindle-shaped
if $ {\sigma}_{\rm P} < {\sigma}_{\rm H} $.

Next, it is especially important that the explicit analytical
expressions~(\ref{eq:j_cond}) enable us to solve easily the inverse problem,
\textit{i.e.}, to derive the electric conductivities in some point of
space~$ (r, \vartheta, \varphi) $ from the components of electric
current~$ j_{cr} $, $ j_{c\vartheta} $, and~$ j_{c\varphi} $ measured in
the same point.
(On the other hand, numerical solutions of the inverse problems are usually
ill-posed and time-consuming.)
So, resolving the set of equations~(\ref{eq:j_cond}) with respect
to~$ {\sigma}_0 $, $ {\sigma}_{\rm P} $, and $ {\sigma}_{\rm H} $, we get:
\begin{eqnarray}
\!\!\!\!\!\!\!\!\! {\sigma}_{\rm H} \!\! & = & \!\!
  \frac{1}{4} \, \frac{c}{B_0 u_0} \, \frac{R}{r} \,
  \frac{ 3 j_r \sin(2 \vartheta)
  + j_{\vartheta} \big[ 1 + 3 \cos(2 \vartheta) \big] }{
  (1 - \sin\vartheta) \cos\vartheta } ,
\label{eq:sigma_H_derived}
\\[0.5ex]
\!\!\!\!\!\!\!\!\! {\sigma}_{\rm P} \!\! & = & \!\!
  {\sigma}_{\rm H} \, \frac{
  j_{\varphi} \big[ \sin(2 \vartheta) - 2 \cos\vartheta \big] \pm 2D }{
  3 j_r \sin(2 \vartheta) +j_{\vartheta} \big[ 1 + 3 \cos(2 \vartheta) \big]
  } ,
\label{eq:sigma_P_derived}
\\[0.5ex]
\!\!\!\!\!\!\!\!\! {\sigma}_0 \!\! & = & \!\!
  2 {\sigma}_{\rm P} \, \frac{
  \big( j_r \cos\vartheta - j_{\vartheta} \sin\vartheta \big)
  ( \sin \theta - 1) }{ j_r \cos \vartheta ( 1 + 2 \sin\vartheta )
  + j_{\vartheta} ( 2 {\cos}^2{\vartheta} - \sin\vartheta ) } ,
\label{eq:sigma_0_derived}
\end{eqnarray}
where
\begin{eqnarray}
8 D^2 & \!\!\! = \!\! &
j^2_r \big[ 3 \sin(5 \vartheta) + 3 \cos(4 \vartheta) - 3 \sin(3 \vartheta)
\nonumber \\
&& \quad
- \, 6 \sin\vartheta - 3 \big]
\nonumber \\
& \!\!\! + \!\! &
j^2_{\vartheta} \big[ \! - \! 3 \sin(5 \vartheta) - 3 \cos(4 \vartheta)
- 5 \sin(3 \vartheta)
\nonumber \\
&& \quad
+ \, 4 \cos(2 \vartheta) - 2 \sin\vartheta - 1 \big]
\nonumber \\
& \!\!\! + \!\! &
j^2_{\varphi} \big[ \! - \! \cos(4 \vartheta) - 4 \sin(3 \vartheta)
+ 4 \cos(2 \vartheta)
\nonumber \\
&& \quad
- 4 \sin\vartheta + 5 \big]
\nonumber \\
& \!\!\! + \!\! &
j_r j_{\vartheta} \big[ 6 \cos(5 \vartheta) - 6 \sin(4 \vartheta)
+ 2 \cos(3 \vartheta)
\nonumber \\
&& \quad
+ \, 4 \sin(2 \vartheta) - 8 \cos\vartheta \big] \, .
\label{eq:D_def}
\end{eqnarray}
The plus/minus sign in formula~(\ref{eq:sigma_P_derived}) should be chosen
so that the entire expression for~$ {\sigma}_{\rm P} $ to be positive.
Of course, the same analytic expressions~(\ref{eq:j_cond}) can be used
to solve the inverse problem in any other setup, for example, when
the same component of the electric current was measured in three different
points of space, \textit{etc.}

In summary, employing the special spectral properties of the stationary
dynamo-effect differential equation, we developed an efficient method of
analytical treatment of this problem.
The corresponding solutions should be especially valuable for dealing with
the inverse problem, \textit{i.e.}, diagnosing the plasma parameters by
the electric fields and currents measured during its expansion in
the external magnetic field.
In principle, it might be expected that the same generalized spherical
functions as introduced here can be employed in treating a more general,
non-stationary dynamo problem, which was outlined in the end of
Section~\ref{sec:Formulation}.
Unfortunately, this is a much more challenging task, which will require
a lot of additional work.

\subsection*{Acknowledgements}

I am grateful to V.I.~Badin and M.G.~Deminov for drawing my attention to
the problem under consideration as well as to the anonymous referee
for interesting suggestions.

\subsection*{Funding statement}

This work was supported by the ongoing institutional funding.
No additional grants to carry out or direct this particular research were
obtained.

\subsection*{Conflict of interest statement}

The author has stated explicitly that there are no conflicts of interest
in connection with this article.

\subsection*{Artificial Intelligence statement}

The author declares that no Artificial Intelligence tools were used in
preparation of this manuscript, apart from the symbolic calculations
toolbox in MatLab.

\subsection*{Data availability statement}

The software used to generate the results of the present study
(MatLab scripts for the symbolic calculations and plotting
Fig.~\ref{fig:Currents}) can be obtained from the corresponding author
upon a reasonable request.

\end{document}